\documentclass[%
 reprint,
superscriptaddress,
 amsmath,amssymb,
article
]{revtex4-1} 

\usepackage{graphicx}
\usepackage{dcolumn}
\usepackage{bm}
\usepackage{gensymb}

\begin{document}

\preprint{APS/123-QED}

\title{High-throughput search for magnetic and topological order\\ in transition metal oxides}

\author{Nathan C. Frey}
\affiliation{Department of Materials Science \& Engineering, University of Pennsylvania, Philadelphia PA 19103}
\affiliation{Energy Technologies Area, Lawrence Berkeley National Laboratory, Berkeley CA 94720}
\author{Matthew K. Horton}
\affiliation{Energy Technologies Area, Lawrence Berkeley National Laboratory, Berkeley CA 94720}
\affiliation{Department of Materials Science \& Engineering, University of California, Berkeley, Berkeley CA 94720}
\author{Jason M. Munro}
\affiliation{Energy Technologies Area, Lawrence Berkeley National Laboratory, Berkeley CA 94720}
\author{Sinéad M. Griffin}
\affiliation{Department of Physics, University of California, Berkeley, California 94720, USA}
\affiliation{Molecular Foundry, Lawrence Berkeley National Laboratory, Berkeley, California 94720, USA}
\affiliation{Materials Science Division, Lawrence Berkeley National Laboratory, Berkeley, California 94720, USA}
\author{Kristin A. Persson}
\affiliation{Energy Technologies Area, Lawrence Berkeley National Laboratory, Berkeley CA 94720}
\affiliation{Department of Materials Science \& Engineering, University of California, Berkeley, Berkeley CA 94720}
\author{Vivek B. Shenoy}
\email[Corresponding author email: ]{vshenoy@seas.upenn.edu}
\affiliation{Department of Materials Science \& Engineering, University of Pennsylvania, Philadelphia PA 19103}

\date{\today}

\begin{abstract}
The discovery of intrinsic magnetic topological order in $\rm MnBi_2Te_4$ has invigorated the search for materials with coexisting magnetic and topological phases. These multi-order quantum materials are expected to exhibit new topological phases that can be tuned with magnetic fields, but the search for such materials is stymied by difficulties in predicting magnetic structure and stability. Here, we compute over 27,000 unique magnetic orderings for over 3,000 transition metal oxides in the Materials Project database to determine their magnetic ground states and estimate their effective exchange parameters and critical temperatures. We perform a high-throughput band topology analysis of centrosymmetric magnetic materials, calculate topological invariants, and identify 18 new candidate ferromagnetic topological semimetals, axion insulators, and antiferromagnetic topological insulators. To accelerate future efforts, machine learning classifiers are trained to predict both magnetic ground states and magnetic topological order without requiring first-principles calculations.
\end{abstract}

\maketitle


\noindent Materials with coexisting quantum phases offer exciting opportunities for solid-state device applications and exploring new physics emerging from the interplay between effects including topology and magnetism \cite{Tokura2019}. Intrinsic magnetic topological materials enable both explorations of fundamental condensed matter physics and next-generation technologies that rely on topological quantum states. Significant progress has been made on classifying and discovering topological materials \cite{Po2017, Bradlyn2017, Vergniory2019, Zhang2019}. Separate efforts have recently enabled high-throughput classification of magnetic behavior \cite{Chen2019, Horton2019}. However, both the theoretical design and experimental realization of magnetic topological materials are confounded by the inherent difficulties in predicting and controlling magnetic order, often arising from strong electron correlations \cite{Watanabe2018, Ivanov2019, Xu2020}. At the intersection of these two quantum orders the focus has mainly been on taking prototypical topological materials like $\rm (Bi,Sb)_2Te_3$ and introducing magnetic dopants \cite{Chang2013, Checkelsky2014}. This doping approach has a number of drawbacks, namely that dopants are hard to control \cite{Carva2020} and the critical temperature for observing exotic physics is low (below 2 K) \cite{Lee2019}. Important challenges remain in assessing the stability (synthetic accessibility) of proposed materials \cite{Zunger2019} and coupling topological property prediction to magnetism.

Recently, there has been a surge of interest in the field due to the experimental realization of intrinsic magnetic topological phases in the van der Waals material $\rm MnBi_2Te_4$ \cite{Zhang2019, Li2019, Otrokov2019}. The $\rm MnBi_2Te_4$ family offers the first opportunity to possibly access the quantum anomalous Hall (QAH) phase, topological axion states, and Majorana fermions in a single materials platform by controlling the interplay between the magnetic and topological orders \cite{Lee2019}. This demonstration of a true magnetic topological quantum material (MTQM) opens new avenues of research into the modeling, discovery, and characterization of magnetic topological materials. Many opportunities remain; in particular, the realization of new magnetic topological phases \cite{Xu2020} and robust order in ambient conditions.

In this work, we develop and apply workflows to automate the calculation of magnetic exchange parameters, critical temperatures, and topological invariants to enable high-throughput discovery of MTQMs. Building on previous work to determine magnetic ground states with density functional theory (DFT) calculations \cite{Horton2019}, we apply the workflow to a subset of over 3,000 transition metal oxides (TMOs) in the Materials Project database \cite{Jain2013}. We focus on TMOs because of the large range of tunability that has already been demonstrated, both through \textit{ab initio} calculations and molecular beam epitaxial growth of single phase and heterostructured oxide compounds, and the tantalizing possibility of incorporating them with oxide electronics. In fact, several previous works have identified potential topological oxide candidates on a case-by-case basis \cite{Jin2013, Yan2013, Weber2019}, though few have been successfully synthesized or measured to be topological \cite{Zunger2019}. The workflow is used to identify candidate ferromagnetic topological semimetals (FMTSMs), axion insulators, and antiferromagnetic topological insulators (AFTIs), as well as layered magnetic and topological materials. Moreover, the computed magnetic orderings and a recently published data set of predicted magnetic topological materials \cite{Xu2020} are used to train machine learning (ML) classifiers to predict magnetic ground states and magnetic topological order, which may be used to accelerate exploration of the remaining 31,000+ magnetic materials in the Materials Project. Future work will extend this modular workflow to explore diverse phenomena including other ferroic orders, exotic topological phases, and new materials systems.

\bigskip
\noindent \textbf{Results}

\begin{figure}[htbp]
\includegraphics[width=\columnwidth]{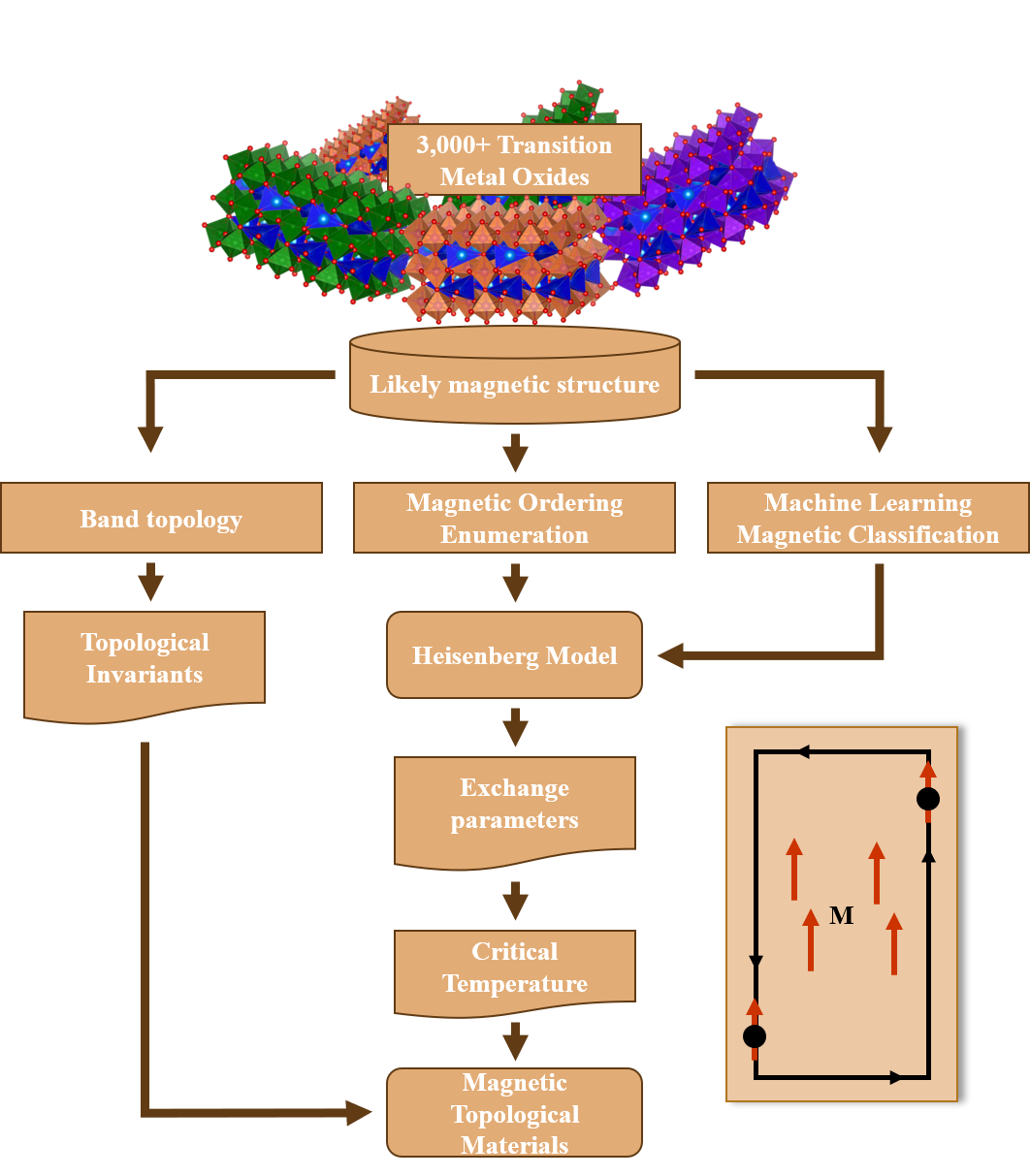}
\caption{\label{fig:fig1} Workflow diagram for high-throughput computation of magnetic ordering, exchange parameters, and topological invariants.}
\end{figure}

\noindent \textbf{Calculating magnetic and topological order.} The workflow presented here is graphically summarized in Fig. \ref{fig:fig1}. For any candidate magnetic material, the method previously developed by some of the coauthors \cite{Horton2019} is used to generate likely collinear magnetic configurations based on symmetry considerations. Exhaustive DFT calculations are performed to compute the energies of each magnetic ordering and determine the ground state. Alternatively, the machine learning classifier discussed below can be used to predict the ground state ordering based solely on structural and elemental data to accelerate the ground state classification. The low energy orderings are then mapped to the Heisenberg model for classical spins, $\vec{S}$:

\begin{equation}
H=-\sum_{i, j} J_{ij} \vec{S_i} \cdot \vec{S_j}.
\end{equation}

Solving the resulting system of equations yields the exchange parameters, $J_{ij}$. The computed exchange parameters and magnetic moments provide all the necessary inputs to obtain the critical temperature through Monte Carlo simulations. The crystal is represented as a structure graph (using the $NetworkX$ package \cite{Hagberg2008}) where nodes represent atoms and edges represent exchange interactions. The entire analysis has been implemented in the \textit{pymatgen} \cite{Ong2013} code and an automated magnetism workflow is available in \textit{atomate} \cite{Mathew2017}. Monte Carlo calculations are enabled in the workflow by interfacing with the VAMPIRE atomistic simulations package \cite{Evans2014}. It should be noted that this method is only applicable for systems that are well described by the classical Heisenberg model, that is, systems with \textit{localized magnetic moments} and reasonably high Curie or Néel temperatures ($ T_{C/N} > 30$ K), such that quantum effects can be neglected.

The second branch of the workflow diagnoses band topology. Topological invariants are determined using the $vasp2trace$ \cite{Vergniory2019} and $irvsp$ \cite{Gao2020} codes to compute irreducible representations of electronic states, as well as the hybrid Wannier function method in $Z2Pack$ \cite{Gresch2017}. Automated workflows to calculate topological invariants are implemented in the \textit{Python Topological Materials (pytopomat)} code \cite{pytopomat2020}. By coordinating workflows, we are able to discover materials with coexisting quantum orders, like magnetic topological insulators, in a high-throughput context. The schematic in Fig. \ref{fig:fig1} shows one such example: a magnetic system exhibiting the quantum anomalous Hall effect.

\bigskip
\noindent \textbf{Transition metal oxide database.} We restrict our search to the family of transition metal oxides (TMOs), which has the advantages of encompassing thousands of candidate magnetic materials and having standardized Hubbard \textit{U} values based on experimental enthalpies of formation \cite{Jain2011}. A subset of 3,153 TMOs were considered, encompassing over 27,000 computed magnetic orderings, with any combination of Ti, V, Cr, Mn, Fe, Co, Ni, Cu, O, and any other non-\emph{f}-block elements. Importantly, only stable and metastable phases within 200 meV of the convex hull are included in our database. For each TMO, up to 16 likely magnetic orderings were generated, yielding a total of 923 ferromagnetic (FM) and 2,230 antiferromagnetic (AFM) ground states. For simplicity, ferrimagnetic (FiM) ground states were called AFM if they have an anti-parallel spin configuration with a net magnetic moment less than 0.1 $\mu_B$/cell, and FM if the net magnetic moment in the system is greater than 0.1 $\mu_B$/cell.

\begin{figure}[htbp]  
\includegraphics[width=\columnwidth]{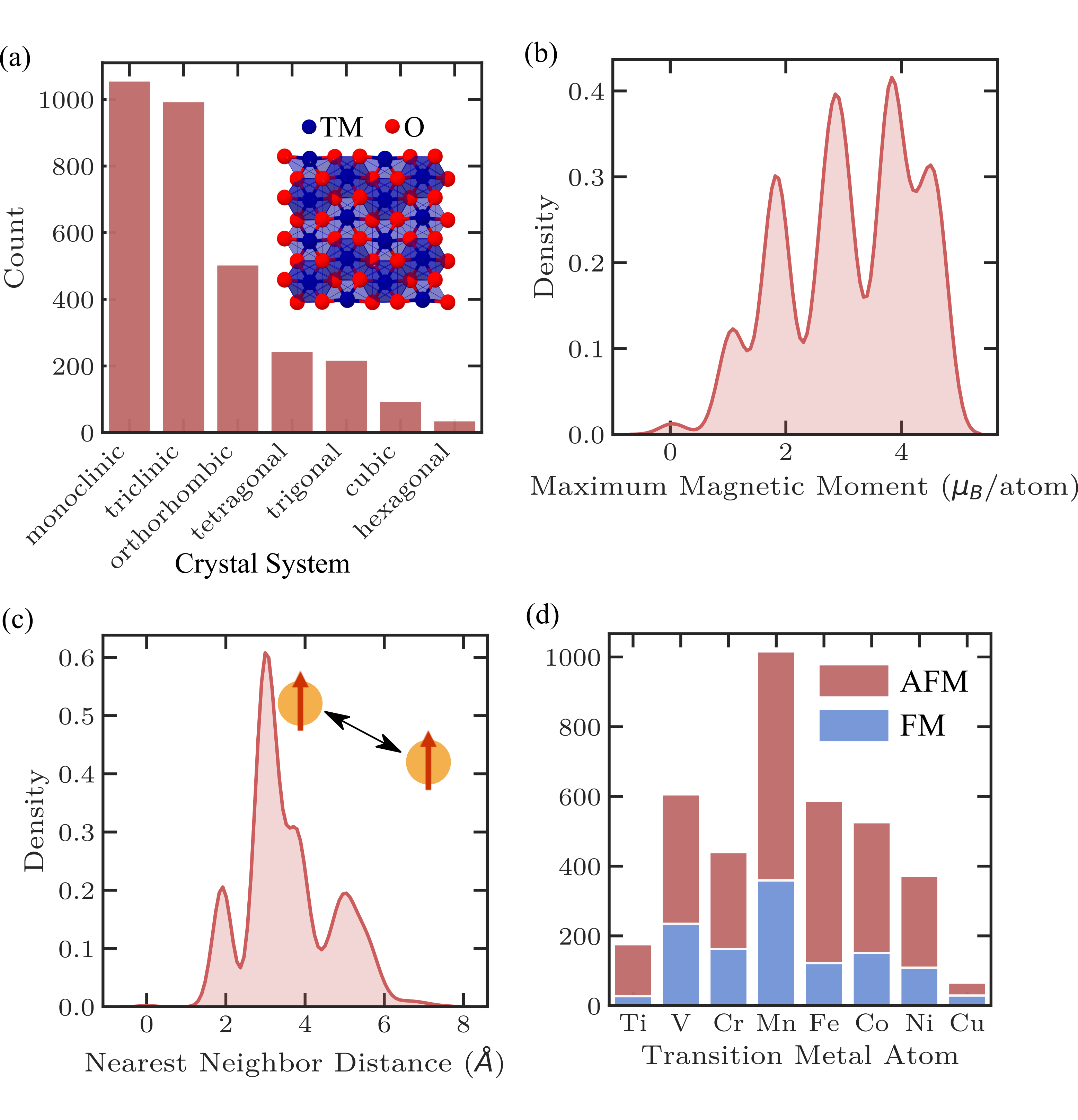}  
\caption{\label{fig:fig2}Survey of magnetic transition metal oxides in the database. (a) Histogram of crystal systems. (b) Maximum magnetic moments in each system. Clustering is observed around integer values. (c) Average nearest neighbor distance between magnetic ions. (d) Occurrence of 3$d$ block transition metal atoms across FM and AFM systems.}
\end{figure}

A statistical summary of the data set is presented in Fig. \ref{fig:fig2}. Fig. \ref{fig:fig2}(a) is a histogram of the crystal systems contained in the data. All seven crystal systems are represented, with monoclinic being the most prevalent and hexagonal the least. Similarly, there are a variety of space groups, compositions, and symmetries present in the materials considered. There are compounds with one, two, three, or more magnetic sublattices. As expected for TMOs, most compounds have an average coordination number of four or six. Considering the computed ground states, there is a large range of maximum magnetic moments per atom, with clustering observed around the integer values of 1, 2, 3, and 4 $\mu_B$/atom (Fig. \ref{fig:fig2}(b)). The histogram of average nearest neighbor distance between two TM atoms in a compound is shown in Fig. \ref{fig:fig2}(c). There is again a large range of values, from 2 to 6 $\rm \AA$, with a peak around 3 $\rm \AA$. We also show the relative occurrence of 3$d$ block transition metal species across FM and AFM compounds in Fig. \ref{fig:fig2}(d). Mn is the most common transition metal, occurring in 1016 compounds in the database, while Cu is the least prevalent, occurring in fewer than 100 compounds.

\begin{figure}[htbp] 
\includegraphics[width=\columnwidth]{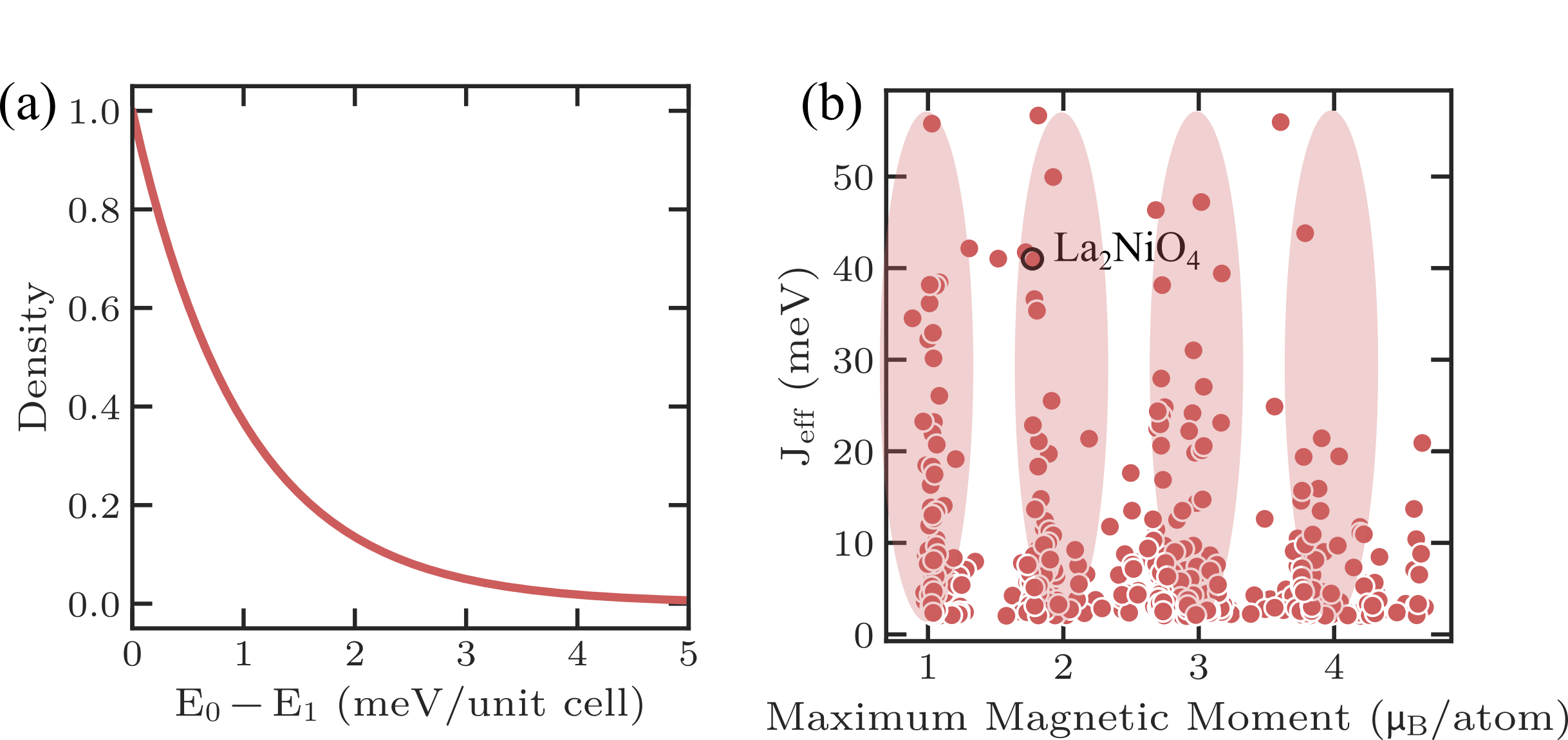}
\caption{\label{fig:fig3}Low-energy ordering and effective exchange interactions. (a) Energy splitting between ground state and first excited state. (b) $J_{eff}$ versus maximum magnetic moment in the unit cell.}
\end{figure}

A wealth of information is available in the computed higher energy orderings as well. We define the energy gap, $\Delta E = E_0 - E_1$, where $E_0$ is the ground state energy and $E_1$ is the energy of the first excited state. When the low-energy orderings are successfully found, $\Delta E$ quantifies the robustness of the ground state ordering. The plot of $\Delta E$ in Fig. \ref{fig:fig3}(a) shows the heavy-tailed distribution of the energy gaps. Over 600 compounds exhibit $\Delta E < 0.5$ meV/unit cell and may have correspondingly small $J$ and $T_{C/N}$ values. An effective $J$ parameter can be estimated from the energy gap as $|J_{eff}|=\Delta E / (NS^2)$, where $N$ is the number of magnetic atoms in the unit cell and $S$ is the magnitude of the average magnetic moment. From this crude estimate, the transition temperature is given in the mean field approximation as $T_{C/N}^{MFT}=2J_{eff}/(3k_B)$, where $k_B$ is Boltzmann's constant. The plot in Fig. \ref{fig:fig3}(b) shows the $J_{eff}$ values, more or less clustered by the integer values of the maximum magnetic moment (indicated by the light red ovals) in each material. One representative high $J_{eff}$ material is ${\rm La_2NiO_4}$ (41.7 meV), which has an AFM ground state and an estimated $T_N$ of 323 K (measured value of 335 K \cite{Perkins1995}).

We briefly highlight some promising material candidates that may be reduced into a two-dimensional (2D) form \cite{Frey2019} with possible access to intriguing low-dimensional magnetic properties \cite{Kalantar-zadeh2016, Bandyopadhyay2019}, and materials with strong spin-orbit coupling (SOC). We apply the method from Ref. \cite{Gorai2016} to identify potentially layered TMOs, which yields 105 candidates. There are also 66 TMOs that contain either Bi or Hg and are therefore expected to exhibit strong SOC. At the intersection, we find three Bi-containing layered magnetic materials, ${\rm Ba_2Mn_2Bi_2O}, {\rm CoBiO_3}$ and ${\rm CrBiO_4}$. These three materials may be of particular interest as tunable 2D magnets with strong SOC-induced magnetic anisotropy \cite{Frey2018}.

The primary computational burden in generating this data set is calculating the relaxed geometries and energies of all likely magnetic orderings, as we have no \emph{a priori} way of determining the magnetic ground states. Further, it is useful to compute the spectrum of low-energy magnetic orderings to estimate the strength of exchange couplings, thereby determining the nature of magnetic interactions and critical temperatures. For simple compounds with small unit cells and a single type of magnetic ion, it is relatively easy to determine the ground state and only a few ($< 4$) orderings need to be computed. However, there is a long tail to the number of orderings required for complex structures that may have many highly symmetric AFM orderings \cite{Horton2019}. For the TMO data set, nine orderings per compound are required, on average, to find the ground state. Due to the computational cost, we are limited to collinear magnetic orderings in this combinatorial approach, although this work is an important first step towards determining the noncollinear orderings. It is highly desirable to augment these laborious DFT calculations with computationally inexpensive, physics-informed models that can predict magnetic behavior.  

\bigskip
\noindent \textbf{Magnetic ordering machine learning classifier.} The size of the TMO data set and the number of easily available, physically relevant descriptors suggests that a physics-informed machine learning classifier may be able to predict magnetic ground states. Our goal is to use features based purely on structural and compositional information, \emph{without any DFT calculations}, to predict magnetic orderings and prioritize calculations. With the \emph{matminer} \cite{Ward2018} package, we have access to thousands of descriptors that are potentially correlated with magnetic ordering. Drawing on physical and chemical intuition, this list was reduced to $\sim100$ descriptors that are likely indicators of magnetic ordering, e.g. elemental \emph{d} orbital filling, electronegativity, and tabulated atomic magnetic moments. We also generate additional features more specific to magnetic compounds, including the average nearest neighbor distance between TM atoms, TM-O-TM bond angle information, TM atom coordination number, and the number of magnetic sublattices. We have implemented these features in the `magnetism' module of \emph{pymatgen}. Unsurprisingly, no features have Pearson correlation coefficients larger than 0.3 with respect to ground state ordering. There are no features with strong enough linear correlation to reliably predict magnetic behavior. To further reduce the feature space, we train a minimal model and discard features with extremely low impurity importance and then perform hierarchical clustering based on the Spearman rank correlation, removing a feature from each cluster.

\begin{figure}[htbp]
\includegraphics[width=8.6cm]{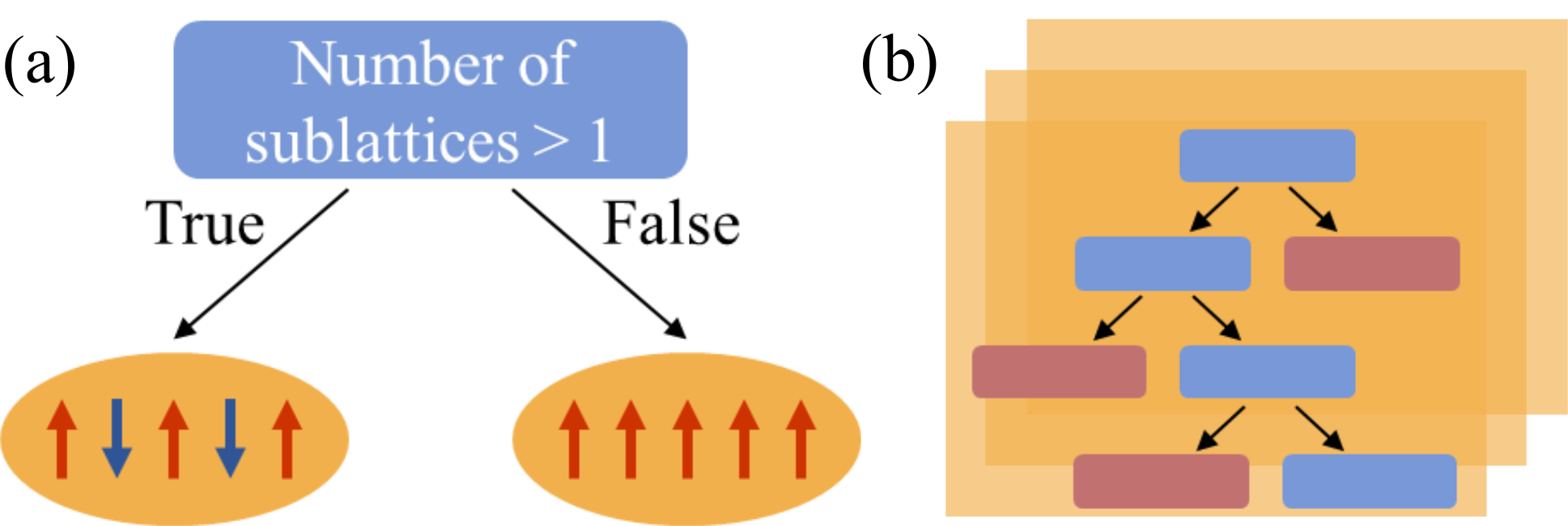}
\caption{\label{fig:fig4}Random forest classifiers for magnetic ground state prediction. (a) An example of a ``leaf" in the decision tree. (b) Graphic representation of a decision tree in the random forest.}
\end{figure}

Next, using the reduced feature set, we construct an ensemble of machine learning classifiers to predict the magnetic behavior. For simplicity and interpretability, a random forest classifier \cite{Breiman2001} was used, although other techniques like Adaptive Boosting and Extra Trees perform similarly. The random forest is an ensemble of decision trees made up of ``leaves" like the one shown schematically in Fig. \ref{fig:fig4}(a). For each feature, the tree splits the data set to enable classification. In the illustration in Fig. \ref{fig:fig4}(a), the simplified split illustrates that samples with more than one magnetic sublattice are more likely to be AFM than FM. To capture the complexity of the data, a full decision tree is more fleshed out, like the one shown in Fig. \ref{fig:fig4}(b). The random forest is an ensemble of many such trees, where the predictions of uncorrelated trees are averaged over to reduce overfitting. $10\%$ of the data was held as a test set, and five-fold cross-validation was used to tune the model hyperparameters. Because of the class imbalance between FM and AFM ground states (30\% of compounds are FM), the FM compounds are synthetically oversampled using SMOTE \cite{Chawla2002}. This leads to good performance in five-fold cross-validation, as seen in the mean and median $F_1$ scores of 0.85 for both FM and AFM classes (see the Supplemental Material (SM) for details). The trained model achieved an $F_1$ score of 0.85 (0.59) for AFM (FM) compounds on the test set, suggesting that the synthetic oversampling results in difficulties generalizing to new FM compounds, while AFM systems are well characterized.

The success of the classifier allows us to reexamine the input features and use the model feature importances to identify nontrivial predictors of magnetic behavior. Whereas the Gini impurity-based feature importance somewhat misleadingly shows equal contributions from many features, here we use the permutation importance, which avoids bias towards numerical and high cardinality features \cite{Breiman2001}. The most important features for classification are shown in Fig. S1. By far the most important descriptor is the number of magnetic sublattices. Other features relate to space group symmetry, \emph{d} electron counts, coordination number, and distances between TM atoms; features we expect to describe magnetism. Another important descriptor is the sine Coulomb matrix, which is a vectorized representation of the crystal structure that has been introduced and used in previous studies to predict atomization energies of organic molecules \cite{Rupp2012} and formation energies of crystals \cite{Faber2015}. Finally, the structural complexity \cite{Krivovichev2013} is observed to be $47\%$ higher on average for AFM compounds than FM. The AFM systems exhibit an average structural complexity that is 17 bits/unit cell higher than the FM systems. This simple metric might indicate that more structurally complex materials are more likely to favor the more complex AFM orderings, rather than simple FM configurations. This could be related to the most important descriptor, which is also a metric of magnetic lattice complexity. Surprisingly, these complexity metrics along with simple TM-TM atom distances and the sine Coulomb matrix are much better predictors of magnetic ordering than bond angle information, as might be expected from the Goodenough-Kanamori rules. It is possible that more sophisticated features may do a better job at capturing the superexchange mechanisms that govern the magnetic behavior.

It is clear that models like the one presented here are not guaranteed to generalize beyond the material types that comprise the training data \cite{Meredig2018}. However, we expect that the physical insights related to feature engineering, as well as the tested methods, will be of use in future studies. Further difficulties will be encountered when constructing machine learning models for critical temperature prediction, which is inherently a problem of outlier-detection. Fortunately, this work provides both a set of promising materials to consider for further study and the framework to automate evaluation of exchange parameters and critical temperatures. 

\begin{figure*}[htbp]
\includegraphics[width=16cm]{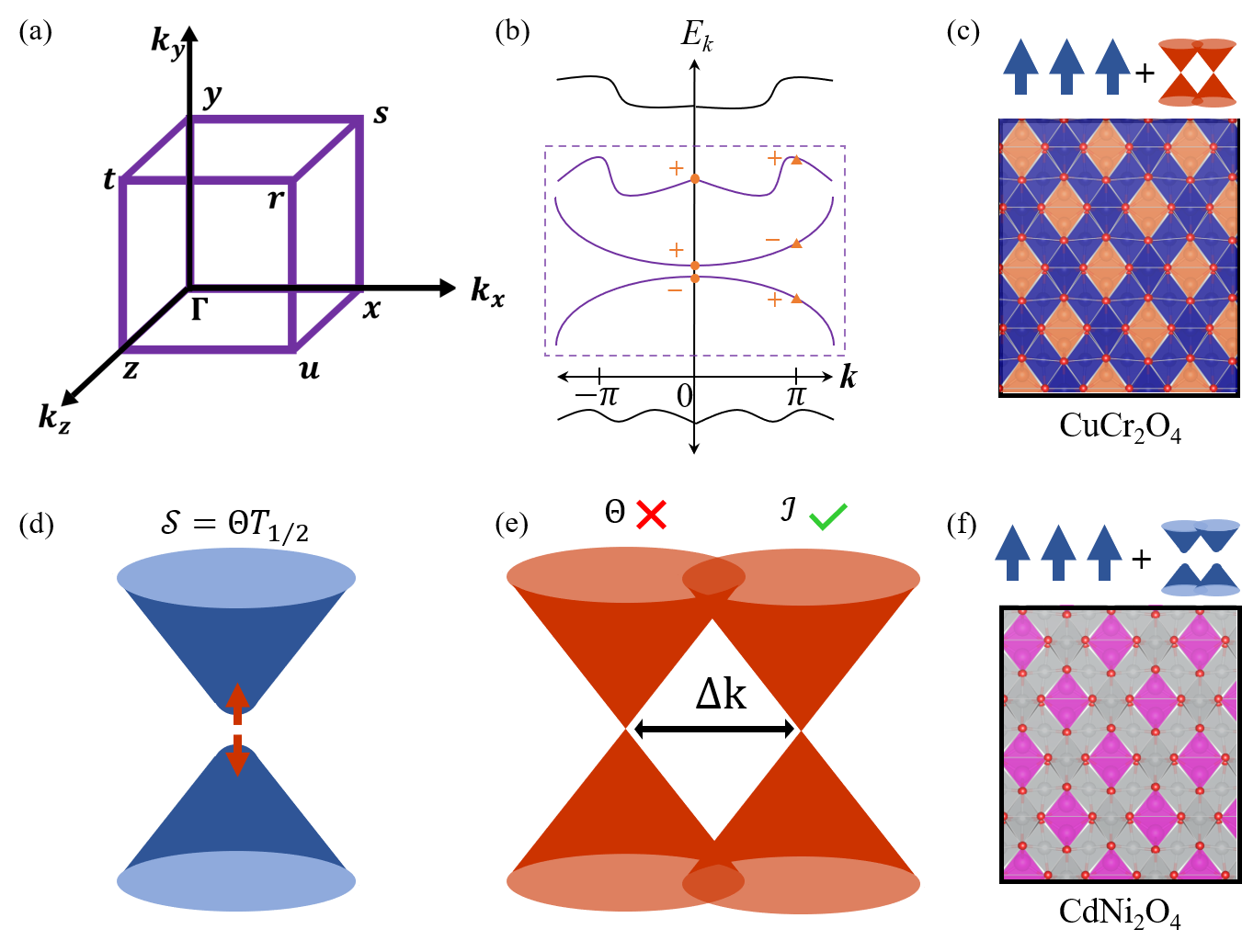}
\caption{\label{fig:fig5}Magnetic topological materials. (a) Time-reversal invariant momenta in the Brillouin zone. (b) Schematic of parity eigenvalues of occupied bands at TRIM points. (c) The candidate ferromagnetic topological semimetal, spinel ${\rm CuCr_2O_4}$. (d) Schematic of a Dirac cone in an antiferromagnetic topological insulator with $\mathcal{S}$ symmetry. (e) Schematic of Weyl cones in a ferromagnetic topological semimetal without time-reversal and with inversion symmetry. (f) The candidate ferromagnetic axion insulator, spinel ${\rm CdNi_2O_4}$.}
\end{figure*}

\bigskip
\noindent \textbf{Magnetic topological material identification.} Finally, we discuss the search for nontrivial band topology in the magnetic TMOs. The zoo of available topological order is ever expanding; here, we simplify our search by considering classes of centrosymmetric magnetic topological materials that can be readily classified with high-throughput calculations of topological indices. We consider antiferromagnetic topological insulators (AFTIs), ferromagnetic topological semimetals (FMTSMs), and ferromagnetic axion insulators. In the first case, we consider materials that exhibit an AFM ground state that breaks both time-reversal ($\Theta$) and a primitive-lattice translational symmetry ($T_{1/2}$), but is invariant under the combination $\mathcal{S}=\Theta T_{1/2}$. The preserved $\mathcal{S}$ symmetry allows for the definition of a $\mathbb{Z}_2$ topological invariant \cite{Mong2010} that lends itself to high-throughput evaluation. For ferromagnets, we consider FM ground states that break $\Theta$ symmetry but preserve inversion symmetry ($\mathcal{I}$), because their band topology can be determined by the parity eigenvalues of occupied bands at the eight time-reversal invariant momenta (TRIM) in the Brillouin zone (BZ) (Fig. \ref{fig:fig5}(a,b)). Specifically, we restrict our search to ferromagnets with centrosymmetric tetragonal structures, where ideal Weyl semimetal (WSM) features may appear and where the magnetization direction can tune the band topology \cite{Jin2017}. These filters greatly simplify the screening, but recent work suggests that over 30\% of non-magnetic \cite{Vergniory2019, Claussen2019} and magnetic \cite{Xu2020} materials exhibit nontrivial topology, so there are almost certainly many more interesting MTQMs to uncover in the TMO data set than we have considered here.

We classify potential $\mathbb{Z}_2$ phases in AFM systems by the set of indices
\begin{subequations}
\begin{align}
&\mathbb{Z}_2 = (v;v_{x} v_{y} v_{z}) ,
\\
&v = \Delta(k_i = 0) + \Delta(k_i = 1/2) \; \rm mod \: 2,
\\
&v_i = \Delta(k_i = 1/2),
\end{align}
\end{subequations}
where $\Delta(k_i)$ is the 2D topological invariant on the time-reversal invariant plane $k_i$ in the BZ, and $k_i$ is in reduced coordinates. If $v=1$, the system is a strong topological insulator, while a system with $v=0$ and $v_i=1$ for any $v_i$ is a weak topological insulator \cite{Fu2007, Gresch2017}. FMTSMs are classified by the strong topological index $\mathbb{Z}_4$ in terms of parity eigenvalues \cite{Watanabe2018, Turner2012, Xu2020}, defined as
\begin{equation}
\begin{split}
\mathbb{Z}_4=\sum_{\alpha=1}^{8}\sum_{n=1}^{n_{occ}} \frac{1+\xi_n(\Lambda_\alpha)}{2} \; \rm mod \: 4 ,
\end{split}
\end{equation}
where $\Lambda_\alpha$ are the eight TRIM points, $n$ is the band index, $n_{occ}$ is the number of occupied bands, and $\xi_n(\Lambda_\alpha)$ is the parity eigenvalue $(\pm 1)$ of the $n$-th band at $\Lambda_\alpha$. $\mathbb{Z}_4 = 1,3$ indicates a WSM phase with an odd number of Weyl points in half of the BZ, while $\mathbb{Z}_4=2$ indicates an axion insulator phase with a quantized topological magnetoelectric response \cite{Liu2020}, or a WSM phase with an even number of Weyl points. $\mathbb{Z}_4=0$ corresponds to a topologically trivial phase.

\begin{table}[htbp]
\caption{\label{tab:table1}Candidate tetragonal ferromagnetic topological semimetals and axion insulators. Theoretical materials that have not yet been experimentally synthesized are labeled with a $\dagger$.}
\begin{ruledtabular}
\begin{tabular}{lcccc}
\textrm{Material}&
\multicolumn{1}{p{1.5cm}}{\centering Space \\ group}&
\multicolumn{1}{p{2cm}}{\centering Materials Project \\ ID}&
\multicolumn{1}{p{1.5cm}}{\centering Energy above hull \\ (meV/atom)}&
\textrm{$\mathbb{Z}_4$}\\
\colrule
$\rm CuFe_2O_4$ & $I4_1/amd$ & N/A & N/A & 1\\
$\rm CrO_2$ & $P4_2/mnm$ & mp-19177 & 63 & 3\\
$\rm Sr_3CaFe_4O_{12}^\dagger$ & $P4/mmm$ & mp-1076424 & 14 & 3\\  
$\rm Mn_3O_4F_2^\dagger$ & $P4_2/mnm$ & mp-780777 & 76 & 2\\
$\rm Sr_2La_2Mn_4O_{11}$ & $I4/mmm$ & mp-1218776 & 65 &
 2\\
$\rm Mn_2PO_5^\dagger$ & $I4_1/amd$ & mp-754106 & 27 & 1\\
$\rm Sr_5Mn_5O_{13}$ & $P4/m$ & mp-603888 & 2 & 2\\
$\rm CaV_2O_4^\dagger$ & $I4_1/amd$ & mvc-10887 & 31 & 2\\
$\rm CdNi_2O_4^\dagger$ & $I4_1/amd$ & mp-756341 & 0 & 2\\
$\rm Cr_2TeO_6$ & $P4_2/mnm$ & mp-21355 & 0 & 2\\
$\rm CuCr_2O_4$ & $I4_1/amd$ & mp-1103973 & 13 & 1\\
$\rm LiNiO_2^\dagger$ & $I4_1/amd$ & mp-770635 & 17 & 2\\
$\rm VMg_2O_4^\dagger$ & $I4_1/amd$ & N/A & N/A & 2\\
\end{tabular}
\end{ruledtabular}
\end{table}

The TMO database was screened for materials with FM ground states, a tetragonal crystal structure, and inversion symmetry, resulting in 27 candidates. By computing the $\mathbb{Z}_4$ indices for these materials, we identify eight materials with $\mathbb{Z}_4=2$, indicating either a WSM phase with an even number of Weyl points in half of the BZ, or an axion insulator phase. Five materials have an odd number of Weyl points in half of the BZ, with $\mathbb{Z}_4=1,3$. The candidate FMTSMs and axion insulators and their respective $\mathbb{Z}_4$ indices are listed in Table \ref{tab:table1}. We also give the unique identifiers for the Materials Project database entries and the calculated energy above the convex hull. Here, we highlight the candidate FMTSM $\rm CuCr_2O_4$ (Fig. \ref{fig:fig5}c). $\rm CuCr_2O_4$ has an FM ground state and $\mathbb{Z}_4=1$. $\rm CuCr_2O_4$ is a hausmannite-like spinel structure with the tetragonal $I4_1/amd$ space group. Cr atoms bond with O atoms to form $\rm CrO_6$ octahedra that share corners with $\rm CuO_4$ tetrahedra. $\rm Cr^{3+}$ atoms occupy Wyckoff position 8$d$, $\rm Cu^{2+}$ occupy Wyckoff 4$a$, and $\rm O^{2-}$ occupy Wyckoff 16$h$. We also draw special attention to the spinel $\rm CdNi_2O4$ (Fig. \ref{fig:fig5}f), which is predicted to be an FM axion insulator  with $\mathbb{Z}_4=2$ and a bandgap $E_{bg} = 0.125$ eV. This material has not yet been successfully synthesized and represents one of many promising opportunities to grow new magnetic oxides and investigate their topology.

\begin{table}[htbp]
\caption{\label{tab:table2}Candidate antiferromagnetic topological insulators. Theoretical materials that have not yet been experimentally synthesized are labeled with a $\dagger$.}
\begin{ruledtabular}
\begin{tabular}{lcccc}
\textrm{Material}&
\multicolumn{1}{p{1.5cm}}{\centering Space \\ group}&
\multicolumn{1}{p{2cm}}{\centering Materials Project \\ ID}&
\multicolumn{1}{p{1.5cm}}{\centering Energy above hull \\ (meV/atom)}&
\textrm{$\mathbb{Z}_2$}\\
\colrule
$\rm FeMoClO_4$ & $P4/nmm$ & mp-23123 & 6 & (1;100)\\
$\rm MnMoO_4$ & $P2/c$ & mp-19455 & 5 & (1;001)\\
$\rm Ca_2MnO_3^\dagger$ & $I4/mmm$ & mp-1227324 & 27 & (1;000)\\  
$\rm SrV_{3}O_{7}$ & $Pmmn$ & mp-510725 & 3 & (1;010)\\
$\rm Li_2TiVO_4^\dagger$ & $P2/m$ &  N/A & N/A & (1;001)\\
\end{tabular}
\end{ruledtabular}
\end{table}

\begin{figure}[htbp]
\includegraphics[width=8.6cm]{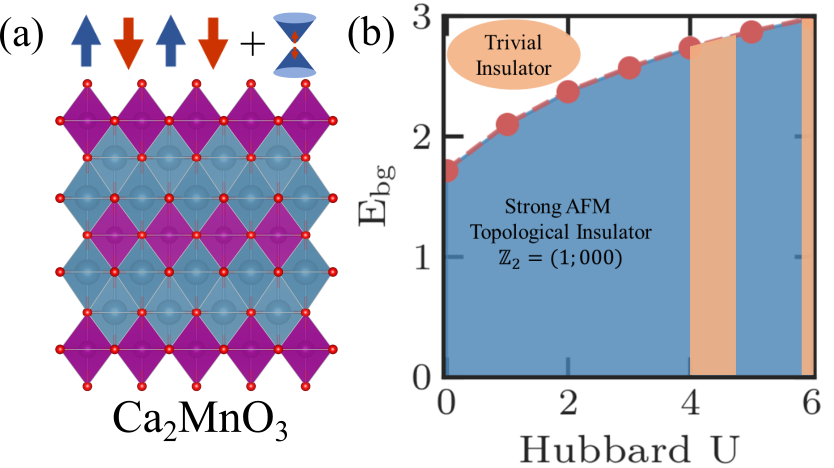}
\caption{\label{fig:fig6}The candidate antiferromagnetic topological insulator, $\rm Ca_2MnO_3$. (a) Crystal structure of $\rm Ca_2MnO_3$  in the tetragonal $I4/mmm$ phase. (b) Phase diagram of bandgap versus the Hubbard $U$ value for Mn showing the dependence of the band topology on the strength of Hubbard interactions.}
\end{figure}

Potential AFTIs were identified by screening the TMO database for AFM ground states with $\mathcal{S}$ symmetry, yielding 298 candidate materials. Of these, 46 are predicted to be layered antiferromagnets by at least one of the methods in Refs. \cite{Gorai2016, Cheon2017, Larsen2019}. These layered systems are of special interest due to their unique and tunable topological and magnetic properties \cite{Kalantar-zadeh2016, Li2019}. Eight additional antiferromagnets with $\mathcal{S}$ symmetry exhibit small bandgaps ($<$ 0.5 eV) and are therefore likely candidates to exhibit band inversion. For each of these 54  materials, the $\mathbb{Z}_2$ invariant is calculated using the hybrid Wannier function method in $Z2Pack$. Four layered AFTIs were identified: $\rm FeMoClO_4$, $\rm MnMoO_4$, $\rm Ca_2MnO_3$, and $\rm SrV_{3}O_{7}$. One small bandgap AFTI was also discovered: monoclinic $\rm Li_2TiVO_4$ in a $P2/m$ phase. These systems and their $\mathbb{Z}_2$ indices are listed in Table \ref{tab:table2}. We highlight the tetragonal $I4/mmm$ phase of $\rm Ca_2MnO_3$ (Fig. \ref{fig:fig6}a), which has a nontrivial $\mathbb{Z}_2 = (1;000)$. It is a caswellsilverite-like structure in which $\rm Ca^{2+}$ ions are bonded with O atoms to form $\rm CaO_6$ octahedra and $\rm Mn^{2+}$ ions bond to form $\rm MnO_6$ octahedra \cite{Ganose2019}. In the primitive cell, Ca atoms occupy Wyckoff position 4$e$, Mn occupies Wyckoff 2$a$, and the O atoms occupy Wyckoff 2$b$ and 4$e$. Because the topology of the AFTI phase is sensitive to the nature of the bandgap and the strength of electron correlations, we plot a phase diagram (Fig. \ref{fig:fig6}b) for $\rm Ca_2MnO_3$ indicating the regions where the system is a strong AFTI or a trivial insulator. We find that the material is a strong AFTI under a wide range of Hubbard $U$ values, although it is predicted to be topologically trivial at $U = 4$ eV and for $U > 6$ eV. Future work will identify the origin of this correlation-dependent change in topological order. 

Importantly, none of the identified candidate MTQMs were considered in previous efforts to screen the Materials Project for topological materials \cite{Vergniory2019}, because the correct magnetic orderings were not available \cite{Horton2019}. We have also highlighted theoretical materials, unique to the Materials Project database, that have not yet been experimentally synthesized and do not have experimental structures reported in the Inorganic Crystal Structure Database (ICSD) \cite{Bergerhoff1983}. Theoretical materials are labeled with a $\dagger$ in Tables \ref{tab:table1} and \ref{tab:table2}. Three materials ($\rm CuFe_2O_4$, $\rm VMg_2O_4$, and $\rm Li_2TiVO_4$) relaxed into new phases not previously included in the Materials Project database after determining the magnetic ground states. Notably, all MTQM candidates are within 100 meV per atom of the convex hull, indicating that all candidate materials are thermodynamically stable or metastable and may be synthesizable \cite{Sun2016}. Additional details on MTQM candidates with ICSD entries and comparisons to experimental measurements of magnetic ordering are given in the SM.

We have extended the machine learning approach discussed above to classify magnetic topological materials from a recently published data set \cite{Xu2020} of 403 magnetic structures containing 130 magnetic topological materials. The random forest model achieves a 0.74 $F_1$ score on topological material classification in five-fold cross-validation, using primarily symmetry- and orbital-based descriptors, requiring no calculations. The details are presented in the SM.

Due to the modularity and interoperability of the workflows developed and applied here, it is straightforward to extend the search to other types of quantum orders. Here, we have provided a high-throughput, relatively coarse-grained method to identify promising MTQMs. The topological structure can be sensitive to the Hubbard $U$ parameter value,  noncollinear magnetic order and the resulting magnetic space group (MSG) determination, and how the strength of SOC compares to the bandgap. Future work will involve detailed studies of candidate materials with the recently introduced Magnetic Topological Quantum Chemistry (MTQC) \cite{Xu2020} formalism, better exchange-correlation functionals (e.g. meta-GGAs like SCAN \cite{Sun2015}) to more accurately compute bandgaps, and careful determination of $U$ values with the linear response approach \cite{Cococcioni2005}.

\bigskip
\noindent \textbf{Discussion} 

\noindent We have developed and applied a high-throughput computational workflow to determine magnetic exchange couplings, critical temperatures, and topological invariants of electronic band structures in magnetic materials. By studying over 3,000 transition metal oxides spanning all crystal systems, nearly all space groups, and a wide range of compositions, we have produced a data set of materials rich in magnetic and topological physics. This enabled the training of a machine learning classifier to predict magnetic ground states and give insight into structural and chemical factors that contribute to magnetic ordering. We extended this machine learning approach to classify topological order in magnetic materials from a recently published data set using only symmetry- and orbital-based descriptors. We identified five promising candidate antiferromagnetic topological insulators (\emph{e.g} tetragonal $\rm Ca_2MnO_3$), including four layered materials, as well as 13 candidate ferromagnetic topological semimetals (spinel $\rm CuCr_2O_4$) and axion insulators (spinel $\rm CdNi_2O_4$).

\bigskip
\noindent \textbf{Methods} 

\noindent DFT calculations were performed with the Vienna Ab Initio Simulation Package (VASP) \cite{Kresse1996, Hobbs2000} and the PBE exchange-correlation functional \cite{Perdew1996}. Standard Materials Project input settings and Hubbard \textit{U} values were used, as described in \cite{Horton2019}. Specifically, values were set for elements Co (3.32 eV), Cr (3.7 eV), Fe (5.3 eV), Mn (3.9 eV), Ni (6.2 eV), and V (3.25 eV) with the rotationally invariant Hubbard correction. These values were determined by fitting to known binary formation enthalpies in transition metal oxides \cite{Jain2011}. Maintaining consistent \emph{U} values with the Materials Project allows for high-throughput calculations and integration within the GGA/GGA+U mixing scheme that enables the construction of phase diagrams. These \emph{U} values and the magnetic ordering workflow were shown to correctly predict non-ferromagnetic ground states in 95\% of 64 benchmark materials with experimentally determined nontrivial magnetic order \cite{Horton2019}. However, it is known that topological phase diagrams for magnetic materials can be strongly dependent on the strength of Hubbard interactions \cite{Xu2020}. Machine learning models were implemented with $scikit-learn$ \cite{scikit-learn}.

\bigskip
\noindent \textbf{Acknowledgements.} N.C.F. was supported by the Department of Defense through the National Defense Science \& Engineering Graduate Fellowship program. V.B.S. acknowledges support from grants EFMA-542879 and CMMI-1727717 from the U.S. National Science Foundation and also W911NF-16-1-0447 from the Army Research Office. S.M.G. and integration with the Materials Project infrastructure was supported by the U.S. Department of Energy, Office of Science, Office of Basic Energy Sciences, Materials Sciences and Engineering Division under Contract No. DE-AC02-05-CH11231 (Materials Project program KC23MP). This research used resources of the National Energy Research Scientific Computing Center (NERSC), a U.S. Department of Energy Office of Science User Facility operated under Contract No. DE-AC02-05CH11231.

\bigskip
\noindent \textbf{Data availability.} The data used in this study are available at https://materialsproject.org. Machine learning models, data used in model training, and an example Jupyter notebook are available on GitHub at https://github.com/ncfrey/magnetic-topological-materials.

\bibliography{TMO}

\end{document}


\preprint{APS/123-QED}

\widetext
\begin{center}
\textbf{\large Supplemental Material for ``High-throughput search for magnetic and topological order in transition metal oxides"}
\end{center}

\author{Nathan C. Frey}
\affiliation{Department of Materials Science \& Engineering, University of Pennsylvania, Philadelphia PA 19103}
\affiliation{Energy Technologies Area, Lawrence Berkeley National Laboratory, Berkeley CA 94720}
\author{Matthew K. Horton}
\affiliation{Energy Technologies Area, Lawrence Berkeley National Laboratory, Berkeley CA 94720}
\affiliation{Department of Materials Science \& Engineering, University of California, Berkeley, Berkeley CA 94720}
\author{Jason M. Munro}
\affiliation{Energy Technologies Area, Lawrence Berkeley National Laboratory, Berkeley CA 94720}
\author{Sinéad M. Griffin}
\affiliation{Department of Physics, University of California, Berkeley, California 94720, USA}
\affiliation{Molecular Foundry, Lawrence Berkeley National Laboratory, Berkeley, California 94720, USA}
\affiliation{Materials Science Division, Lawrence Berkeley National Laboratory, Berkeley, California 94720, USA}
\author{Kristin A. Persson}
\affiliation{Energy Technologies Area, Lawrence Berkeley National Laboratory, Berkeley CA 94720}
\affiliation{Department of Materials Science \& Engineering, University of California, Berkeley, Berkeley CA 94720}
\author{Vivek B. Shenoy}
\email[Corresponding author email: ]{vshenoy@seas.upenn.edu}
\affiliation{Department of Materials Science \& Engineering, University of Pennsylvania, Philadelphia PA 19103}

\maketitle

\beginsupplement

\section{Magnetic ordering classifier interpretation and validation}

\begin{table}[htbp]
\caption{\label{tab:tableS1}$F_1$ score statistics with five-fold cross-validation.}
\begin{ruledtabular}
\begin{tabular}{lccc}
\textrm{Magnetic ordering}&
\textrm{Mean}&
\textrm{Median}&
\textrm{Standard deviation}\\
\colrule
FM & 0.85 & 0.85 & 0.01\\
AFM & 0.85 & 0.85 & 0.01\\
\end{tabular}
\end{ruledtabular}
\end{table}

The $F_1$ score (the harmonic mean of precision, $p$, and recall, $r$) was used to score the model performance:
\begin{subequations}
\begin{align}
&F_1 = 2 \frac{pr}{p+r},
\\
&p = \frac{tp}{tp+fp},
\\
&r = \frac{tp}{tp+fn}.
\end{align}
\end{subequations}
For binary classification with a positive and negative class, $tp$ is the number of true positives, $fp$ is the number of false positives, and $fn$ is the number of false negatives.

The structural complexity, $S$, is given by  
\begin{subequations}
\begin{align}
&S=-\sum_{i=1}^{k} p_{i} \, log_2 \, p_{i} ,
\\
&p_i = m_i/v ,
\end{align}
\label{eqn:all-lines}
\end{subequations}
where $m_i$ is the number of $i$-th inequivalent sites, $k$ is the total number of inequivalent sites, and $v$ is the total number of sites in the unit cell.

\begin{figure}[!htb]
\includegraphics[width=8.6cm]{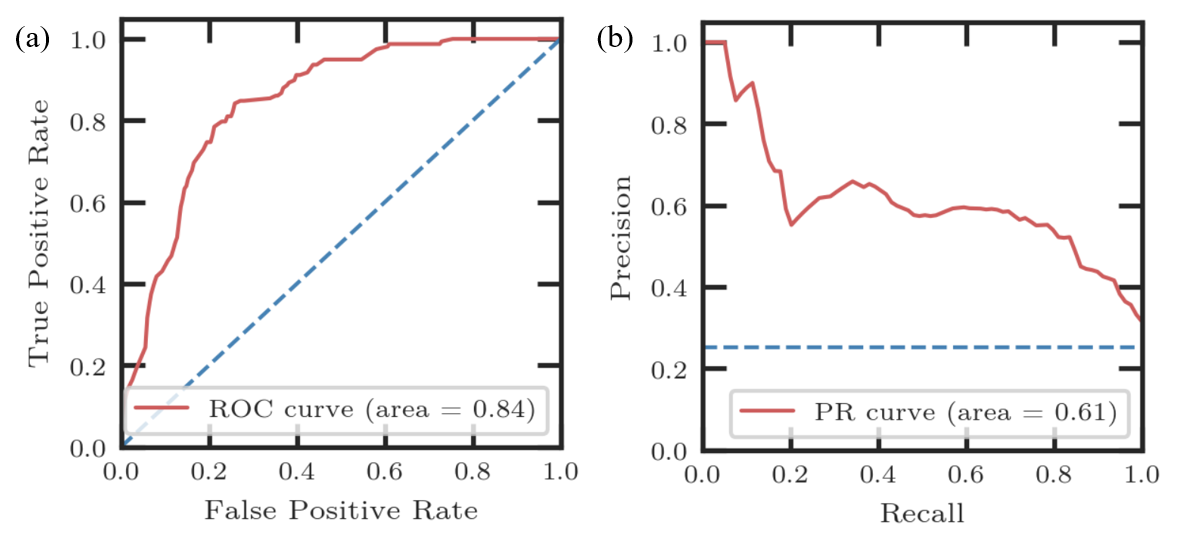}
\caption{\label{fig:figS1}Magnetic ground state classifier performance. (a) Receiver operating characteristic (ROC) curve for test data. Area under the curve (AUC) is 0.84. (b) Precision-recall (PR) curve for test data. AUC is 0.61. Blue dashed lines indicate random guessing.}
\end{figure}

\begin{figure}[!htb]
\includegraphics[width=8.6cm]{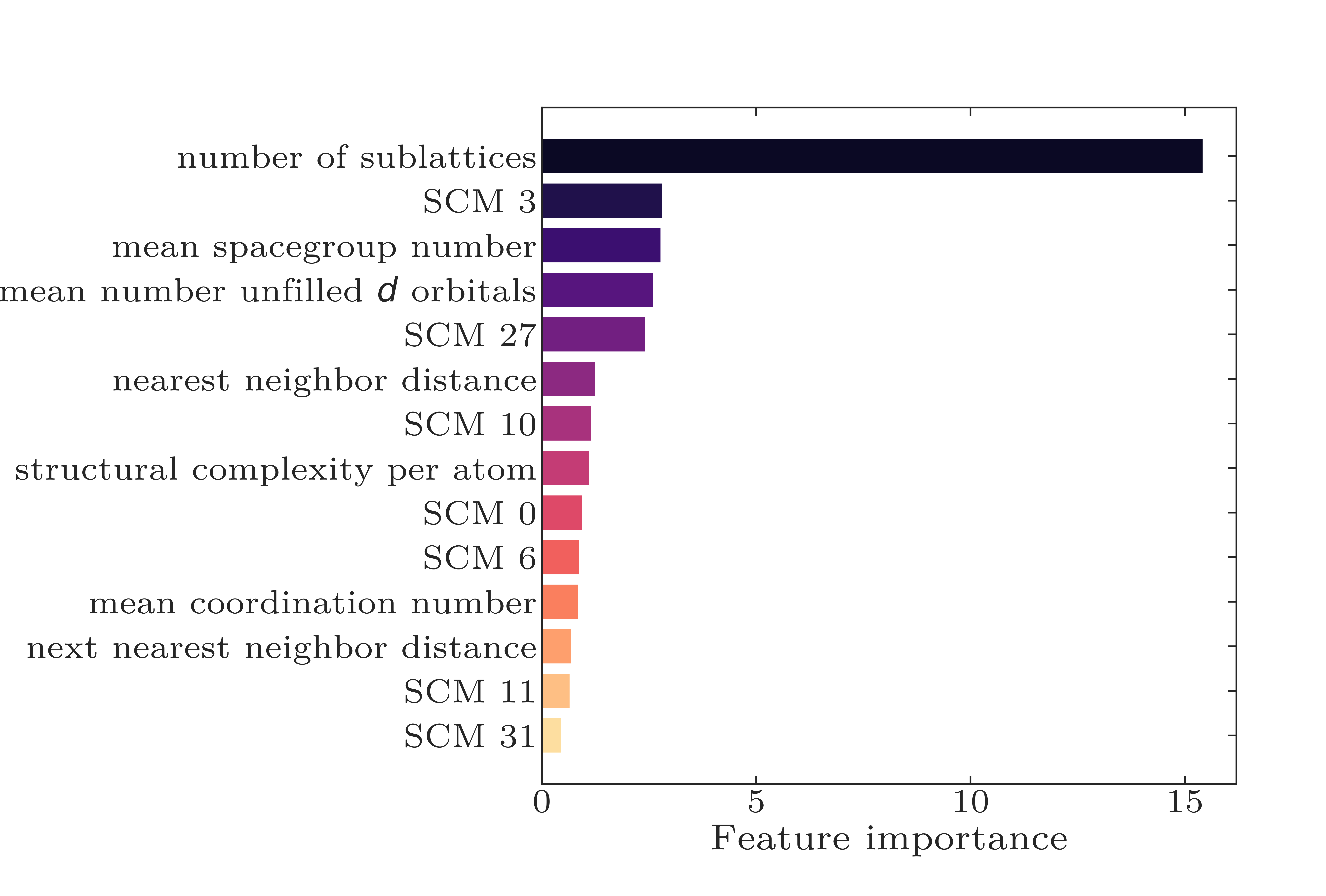}
\caption{\label{fig:figS1}Permutation feature importance graph for the magnetic ordering classifier.}
\end{figure}

\clearpage

\section{Magnetic topological phase machine learning classifier}

\begin{table}[htbp]
\caption{\label{tab:tableS2}$F_1$ score statistics with five-fold cross-validation.}
\begin{ruledtabular}
\begin{tabular}{lccc}
\textrm{Topology}&
\textrm{Mean}&
\textrm{Median}&
\textrm{Standard deviation}\\
\colrule
Nontrivial & 0.74 & 0.73 & 0.04\\
Trivial & 0.75 & 0.75 & 0.03\\
\end{tabular}
\end{ruledtabular}
\end{table}

\begin{figure}[htbp]
\includegraphics[width=8.6cm]{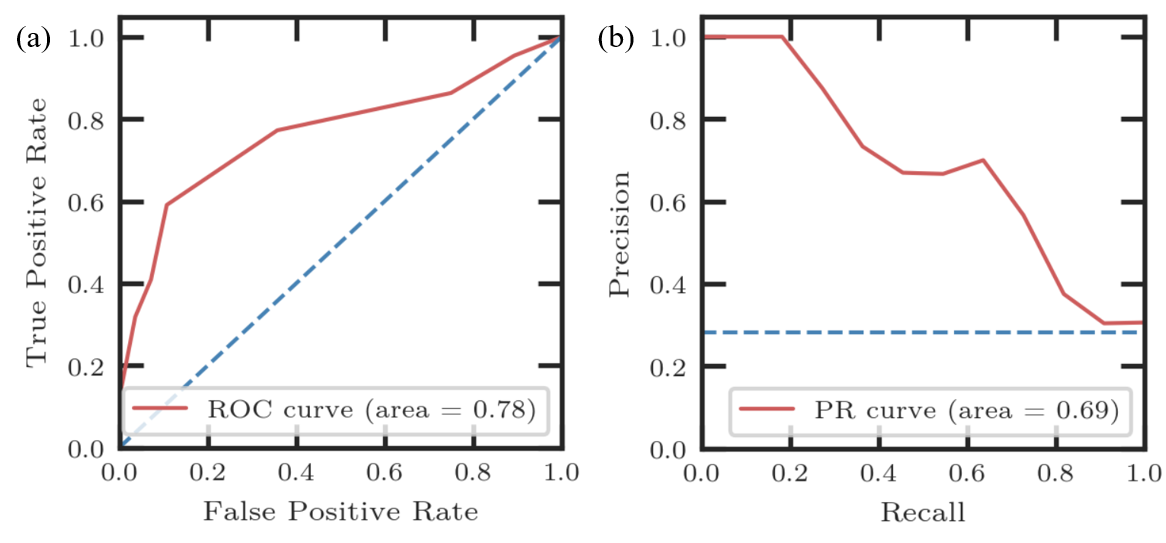}
\caption{\label{fig:figS3}Magnetic topological phase classifier performance. (a) ROC curve for test data. AUC is 0.78. (b) PR curve for test data. AUC is 0.69. Blue dashed lines indicate random guessing.}
\end{figure}

\begin{figure}[htbp]
\includegraphics[width=8.6cm]{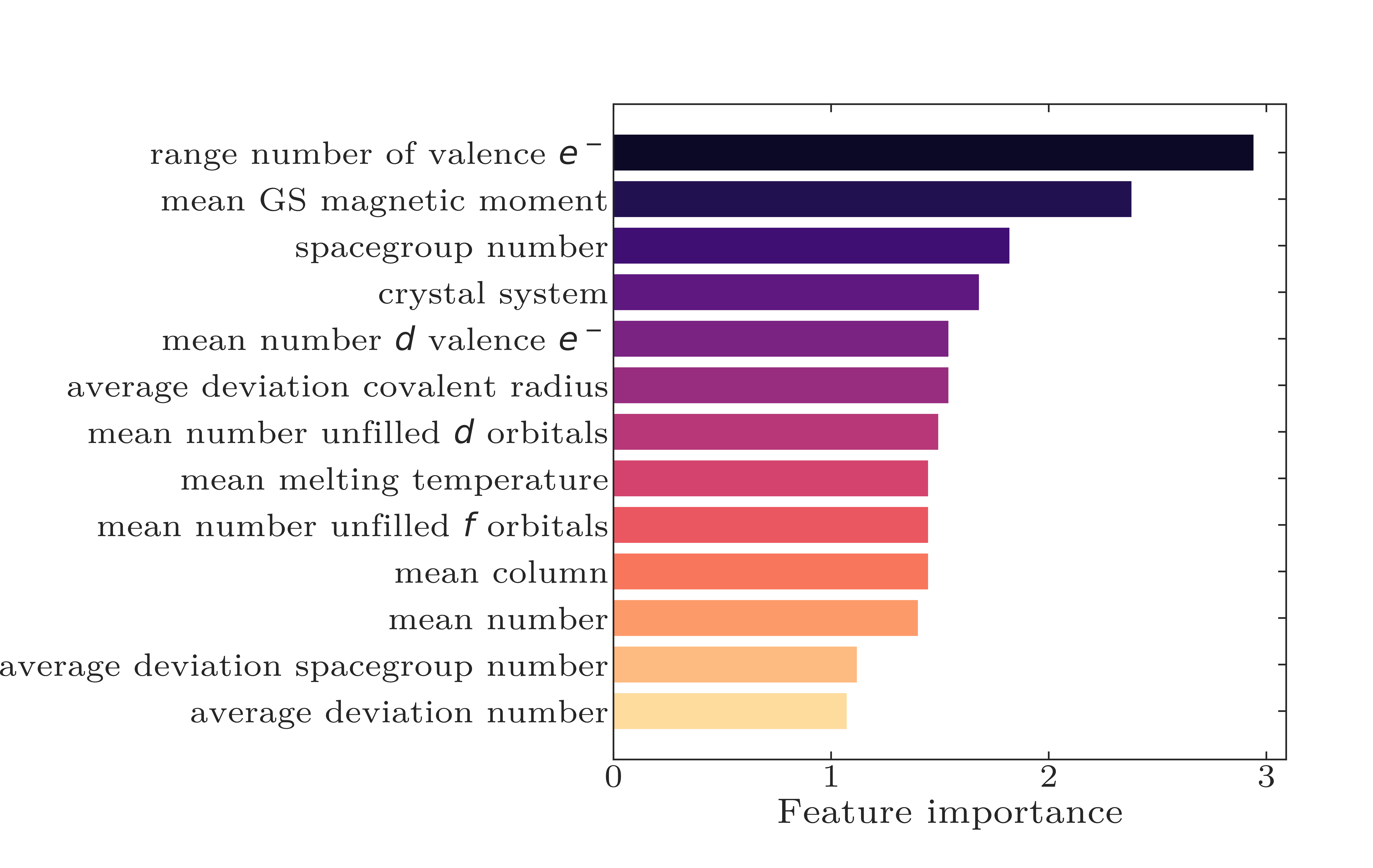}
\caption{\label{fig:figS4}Permutation feature importance graph for the magnetic topological phase classifier.}
\end{figure}

The high-throughput calculations reported in Ref. \cite{Xu2020} started from 707 magnetic structures in the Bilbao Magnetic Material Database (MAGNDATA) \cite{Gallego2016} and converged self-consistent calculations for 403 materials with commensurate structures that are not alloys. Of these, they identified 130 materials ($\sim$32\% of the data) that are magnetic enforced semimetals or topological insulators for at least one tested value of the Hubbard $U$ parameter. We group topological semimetals and insulators into a single class (nontrivial) to avoid a large class imbalance. Using only structural and chemical information, we trained a machine learning classifier to identify magnetic materials that are topologically nontrivial for at least one $U$ value, versus materials that are trivial for all $U$ values. We repeat the model selection and training procedure outlined in the main text. Taking 90\% of the data set and performing five-fold cross-validation, we obtain the $F_1$ score statistics in Table \ref{tab:tableS2}. 10\% of the data is held as a test set and the $F_1$ scores are 0.67 and 0.85 on the topological and trivial materials, respectively. The classifier performance is significantly better than random guessing, although the small data set and complex $U$ dependence make it difficult to achieve high accuracy.

It is interesting to consider the feature importance and model performance and compare these to Ref. \cite{Claussen2019}, in which a gradient boosting model was used to classify non-magnetic topological materials. The reported $F_1$ score for a model with no DFT data was 0.70 for topological insulators (TIs), which is similar to our performance here, despite the 35,000+ materials in the non-magnetic materials data set. This suggests that TIs are difficult to diagnose without detailed electronic structure calculations with SOC, as we might expect because it is challenging to predict band inversion based on stoichiometry and structure. Similarly, we find that features relating to atomic positions are irrelevant for classification performance; instead, symmetry information and ``average orbital character" are decisive, as discussed in Ref. \cite{Claussen2019}. This is reflected in the permutation feature importance graph shown in Fig. \ref{fig:figS4}. The most important features can be grouped generally in terms of symmetry descriptors (space group number, crystal system) and orbital character descriptors (valence electron counts, number of unfilled $d$ and $f$ orbitals, etc.).

\section{Magnetic topological quantum material candidates}

\begin{table}[htbp]
\caption{\label{tab:tableS3}Candidate magnetic topological materials with experimentally determined structures and magnetic orderings.}
\begin{ruledtabular}
\begin{tabular}{lccc}
\textrm{Material}&
\textrm{Space group}&
\textrm{Reported magnetic ground state}\\
\colrule
$\rm FeMoClO_4$ & $P4/nmm$ & AFM \cite{Torardi1984, Torardi1987}\\
$\rm MnMoO_4$ & $P2/c$ & AFM \cite{Clearfield1985,Ehrenberg2006}\\
$\rm SrV_{3}O_{7}$ & $Pmmn$ & AFM \cite{Fudamoto2003}\\
$\rm Cr_2TeO_6$ & $P4_2/mnm$ & AFM \cite{Kunnmann1968}\\
$\rm CuCr_2O_4$ & $I4_1/amd$ & FM \cite{Tamura1994}\\
$\rm CuFe_2O_4$ & $I4_1/amd$ & FM
\cite{Balagurov2013}\\
$\rm CrO_2$ & $P4_2/mnm$ & FM \cite{Cloud1962}\\
\end{tabular}
\end{ruledtabular}
\end{table}

Where possible, experimental measurements of crystal structure and ground state magnetic ordering for candidate magnetic topological materials are listed in Table \ref{tab:tableS3}. The reported orderings agree with the computed ground states in all cases except for $\rm Cr_2TeO_6$, which is interesting because the computed $\mathbb{Z}_4$ index for the FM phase indicates the system may be an axion insulator. $\rm FeMoClO_4$ and $\rm SrV_{3}O_{7}$ are also reported layered materials. 


\bibliography{TMO}